\documentclass[aps,prb,twocolumn,groupedaddress,showpacs,floatfix]{revtex4}
\usepackage{graphicx}
\begin{document}
\pagenumbering{arabic}
\pagestyle{plain}
%
%
%
\title{Internal static electric and magnetic field at the copper cite in a single crystal of the electron-doped high-T$_{c}$ superconductor Pr$_{1.85}$Ce$_{0.15}$CuO$_{4-y}$}  
%
%
\author{Guoqing Wu,$^{1,2}$ F. Zamborszky,$^{2}$ A. P. Reyes,$^{3}$ P. L. Kuhns,$^{3}$ R. L. Greene,$^{4}$ and W. G. Clark$^{2}$}
%
%
\affiliation{$^{1}$College of Physics Science and Technology, Yangzhou University, Yangzhou, Jiangsu 225002, China}
\affiliation{$^{2}$Department of Physics and Astronomy, UCLA, Los Angeles, California 90095, USA}
\affiliation{$^{3}$National High Magnetic Field Laboratory, Florida State University, Tallahassee, Florida 32306, USA}
\affiliation{$^{4}$Department of Physics, University of Maryland, College Park, Maryland 20742, USA}
%
%
%
\date{\today}
\begin{abstract}
     We report $^{63, 65}$Cu-NMR spectroscopy and Knight shift measurements on a single crystal of the electron-doped high-$T_{c}$ superconductor Pr$_{1.85}$Ce$_{0.15}$CuO$_{4-y}$ (PCCO) with an applied magnetic field ($H$) up to 26.42 T. A very small NQR frequency is obtained with the observation of the spectrum, which shows an extremely wide continuous distribution of it that becomes significant narrower below 20 K at $H$ $\parallel$ $c$ where the superconductivity is completely suppressed, indicating a significant change in the charge distribution at the Cu site, while the corresponding changes at $H$ $\perp$ $c$ is negligible when the superconductivity is present or not fully suppressed. The Knight shift and central linewidth are proportional to the applied magnetic field with a high anisotropy. We find that the magnitude of the internal static magnetic field at the copper is dominated by the anisotropic Cu$^{2+}$ 3$d$-orbital contributions, while its weak temperature-dependence is mainly determined by the isotropic contact hyperfine coupling to the paramagnetic Pr$^{3+}$ spins, which also gives rise to the full distribution of the internal static magnetic field at the copper for $H$ $\perp$ $c$. This internal static electric and magnetic field environment at the copper is very different from that in the hole-doped cuprates, and may provide new insight into the understanding of high-$T_{c}$ superconductivity. Other experimental techniques are needed to verify whether the observed significant narrowing of the charge distribution at the Cu site with $H$ $\parallel$ $c$ is caused by the charge ordering (CO) [E. H. da Silva Neto $et ~al.$, to be published in Science] \cite{ehdsn} or a new type of charge modulation.            
\end{abstract}
%
\maketitle
\section{Introduction}
%
     Understanding the mechanism of superconductivity has been an outstanding challenge in physics. The recently advanced technique of resonant inelastic X-ray scattering experiments \cite{dean} in hole-doped cuprate high-$T_{c}$ (where $T_{c}$ is the temperature for the superconducting phase transition) superconductors (HTSCs) La$_{2-x}$Sr$_{x}$CuO$_{4}$ (LSCO) does not support the paring by exchange of magnetic excitations \cite{scalap} as one of the most intensely studied scenarios of high-$T_{c}$ superconductivity, while the proposal \cite{pines1} of spin fluctuations associated with a magnetic interaction between planar quasiparticles (in the CuO$_{2}$-plane) in hole-doped cuprate HTSCs seems to be largely supported. For example, the latter is thought to be responsible for both the anomalous normal state behavior (which strongly deviates from the Fermi liquid theory) \cite{anderson1} and the transition to a superconducting state with an anisotropic orbital $d_{x^2 - y^2}$ wave paring symmetry. \cite{loeser, ding} However, none of them is conclusive. \cite{mann}

     Recent experimental evidence favors a competing scenario in the hole-doped cuprate HTSCs of LSCO, \cite{croft, torch, changj} YBa$_{2}$Cu$_{3}$O$_{y}$ (YBCO) \cite{ghiri, twu, lebo, chang} and Bi$_{2}$Sr$_{2}$CaCu$_{2}$O$_{6+y}$ (BSCCO), \cite{parker, hoffman} which show a competition between superconductivity and other long-range ordered phases, such as a charge-density wave (CDW), \cite{croft, torch, twu, changj} charge ordering (CO), \cite{parker, twu, achkar} and/or even antiferromagnetic (AFM) order, \cite{lake} etc., with the tuning of the applied magnetic field and/or hole-doping level. This interplay between competing phases is also partly observed in the electron-doped cuprate HTSCs Nd$_{2-x}$Ce$_{x}$CuO$_{4-y}$ (NCCO), \cite{hinton} and in the Fe-based high-$T_{c}$ superconductors \cite{yu} as well as in the newly discovered Ti-based \cite{lorenz, doan} superconductors.       

     Unlike the hole-doped cuprate HTSCs, the electron-doped cuprate HTSCs $R_{2-x}$M$_{x}$CuO$_{4-y}$ (RMCO, $R$ = Nd, Pr, Eu, Sm, La; M = Ce, Th) \cite{tokura, jin} show a larger area of antiferromagnetism with no pseudogap phase in the underdoped regime of the phase diagram, \cite{npa} have a substantially lower value of $T_{c}$ (optimal $T_{c}$ $\sim$ 25 K), and have a smaller value of upper critical field ($H_{c2}$) \cite{upperc} than their hole-doped counterparts. These differences suggest \cite {jin, weber} the significance of the antiferromagnetic spin fluctuations which are related to the internal magnetic field environment at the Cu site. On the other hand, in structure they have a slightly longer in-plane Cu-O bond length that might be associated with their lack of apical oxygen for the T$'$-structure of their crystal lattice, in contrast to the T-structure of their hole-doped counterpart in LSCO. \cite{tokura, npa, crr}  This difference in the lattice structure may also have a direct impact on the internal electric and magnetic field environment at the copper which could ultimately determine their spin fluctuations and electron paring. \cite{jin, scalap, pines1, mann, crr} Therefore, it is important to study the internal electric and magnetic field at the copper in the CuO$_{2}$-plane in these materials. Moreover, this local field environment reveals their intrinsic properties, including the sources of the charge and spin dynamics \cite{slichter2} of the Cu-3d conduction electrons.

     NMR (nuclear magnetic resonance) has played a key role in these local field determinations, with intensive studies carried on the hole-doped cuprate HTSCs. \cite{kumagai, warren, barrett, pennington} For example, recent NMR studies of the internal static electric and magnetic field on LSCO \cite{twu1} and YBCO \cite{twu2} have found CO, where the hole-doped charges are modulating around the vortex cores, providing evidence of CO that competes with the superconductivity and suggesting a possible relationship to the electron paring.   

     But there are very few NMR reports on the electron-doped cuprate HTSCs, which were mainly for measurements on powder samples. For example, NMR measurements on powder samples \cite{williams1} of Pr$_{2-x}$Ce$_{x}$CuO$_{4-y}$ (PCCO, $x$ = 0.10, 0.15 and 0.20) by Williams $et ~al.$ were focusing on spin dynamic properties and show conflicting aspects, such as no doping dependence of the $^{63}$Cu-NMR spin-lattice relaxation as a reflection of the internal magnetic field fluctuations at the copper, which are against the widely accepted theoretical predictions. \cite{kobayashi} Thus the intrinsic properties of the electron-doped cuprate HTSCs remain elusive. Since single crystals have lots of advantages over powder samples, it is necessary to examine these properties using single crystals. Recent NMR measurements \cite{mjurk1, mjurk2} on single crystals of PCCO show the effect of doping to the number of hole contents in the Cu 3d and O 2p orbitals and to the $^{63}$Cu-NMR quadrupole splitting frequency. However the important information regarding the distribution of the spectrum satellites and their temperature dependence is still missing.    

     In this paper, we report $^{63, 65}$Cu-NMR spectroscopy and Knight shift measurements on a single crystal of the electron-doped cuprate HTSC PCCO with an applied magnetic field ($H$) up to 26.42 T, at which the superconductivity at $H$ $\parallel$ $c$ is completely suppressed, \cite{upperc} so that the normal state static local field properties at the copper can be evaluated down to low temperatures.      

     Our main results are that a very small NQR frequency $\nu_{Q}$ $\sim$ 2.2 MHz is obtained with the observation of an unusual $^{63, 65}$Cu-NMR spectrum, which shows a very small electric field gradient (EFG) (corresponding to the value of $\nu_{Q}$) and an extremely wide continuous distribution of it ($\Delta \nu_{Q}$ $\sim$ 18 MHz) at the copper site in PCCO. The distribution becomes significantly narrower below 20 K at $H$ $\parallel$ $c$ where the superconductivity is completely suppressed, indicating a significant change in the charge distribution at the Cu site which may be associated with the CO most recently found in the optimally doped NCCO, \cite{ehdsn} while the corresponding changes at $H$ $\perp$ $c$ are negligible when the superconductivity is present or not fully suppressed. The $^{63, 65}$Cu-NMR Knight shift and the central linewidth are proportional to $H$ with a high anisotropy. We find that the magnitude of the internal static magnetic field at the copper is dominated by the anisotropic Cu$^{2+}$ 3$d$-orbital contributions, but its weak temperature ($T$) dependence is mainly determined by the isotropic contact hyperfine coupling to the paramagnetic Pr$^{3+}$ electron spins, which also generates essentially the full distribution of the internal static magnetic field at the copper at $H$ $\perp$ $c$. This internal static electric and magnetic field environment at the copper in the electron-doped cuprate HTSCs is very different from that in their hole-doped counterparts, where there is no evidence of a contribution from ions with a large spin paramagnetic moment.
\section{experimental details}
     The high quality single crystal of PCCO (optimal-doped) was grown with a flux technique \cite{peng,brinkmann} and annealed in argon at 900 $^{\circ}$C for 48 h. The sample size is 1.5 mm $\times$ 1.2 mm $\times$ 35 $\mu$m with a mass of 0.53 mg. The NMR coil was made from 50 $\mu$m diameter silver wire wound with $\sim$ 20 turns, and fixed with epoxy in order to get rid of ``coil disease'' (phonon assisted vibrations). The coil, with the sample in it, was mounted on a small platform that is attached to a goniometer that is rotated about a sample axis. The sample is oriented with the rotation axis in the $ab$-plane and perpendicular to the applied magnetic field $H$ so that the angle ($\Theta$) between the lattice $c$-axis and $H$ can vary as the sample rotates (note, the crystal lattice of PCCO is tetragonal).
    
     The $^{63, 65}$Cu-NMR frequency-swept and field-swept spectra were obtained using standard spin-echo techniques carried out with a spectrometer and probe built at UCLA (W. G. Clark) for field $H$ = 9 T and at the National High Magnetic Field Laboratory (NHMFL) in Florida for field $H$ = 26.42 T, respectively. Since the gyromagnetic ratio $\gamma_{I}$ for the $^{63}$Cu is $^{63}\gamma_{I}$ = 11.285 MHz/T and for the $^{65}$Cu is $^{65}\gamma_{I}$ = 12.089 MHz/T, the frequency $\nu$ for the excitation pulses used for the spectrometer is near $\nu$ $\sim$ $\nu_{0}$ = $\gamma_{I}H$ = $\sim$ 102 (MHz) and $\sim$ 298 (MHz) for the $^{63}$Cu at $H$ = 9 T and $H$ = 26.42 T, respectively, where $\nu_{0}$ is the $^{63}$Cu Larmor frequency in the external field. The corresponding values for the $^{65}$Cu are $\sim$ 109 (MHz) and $\sim$ 319 (MHz) at $H$ = 9 T and $H$ = 26.42 T, respectively.  

     Since the $^{63,65}$Cu-NMR spectrum covers a very wide range in frequency up to $\sim$ 18 MHz (1.5 kG) at all temperatures (and fields), short rf pulses and a wide receiver bandwidth ($\pm$ 1 MHz) were used to record the spin-echo signals. The pulse sequence that optimized the height of the $^{63}$Cu-NMR spin echo (with the central line) used for most of the NMR signal recording was a 0.6 $\mu$s $\pi$/2-pulse ($p_{1}$) [ i.e., rf field $H_{1}$ = 1/(4$^{63}\gamma$$p_{1}$) = 369 G, or 0.42 MHz $^{63}$Cu frequency ] followed by a 1.0 $\mu$s $\pi$-pulse ($p_{2}$) separated by a time interval $\tau$ ($\tau$ $\sim$ 10 $\mu$s) for most of the measurements at both $H$ = 9 T and $H$ = 26.42 T using the same NMR sample coil (note, the optimized pulses for the $^{65}$Cu spin echo is rather similar). For a viable signal-to-noise ratio, each echo signal was averaged 1000 times at 200 K and 64 times at 10 K and lower temperatures at $H$ = 9 T, while at $H$ = 26.42 T the corresponding number of averages used in the measurements is $\sim$ 4 times smaller.

     At $H$ = 9 T, the typical range of the frequency sweep covered 20 MHz (from 95 MHz to 115 MHz) at all temperatures, and it used a frequency step 0.1 MHz for each spin-echo acquisition. In order to maintain a uniform high sensitivity (above 85$\%$), the probe circuit was first tuned to 95.5 MHz, and then retuned manually every 1 MHz (i.e., 10 acquisitions) for the spectrum recording. The frequency-swept spectra were analyzed with the frequency-shifted and -summed (FrSS) Fourier transform processing. \cite{wgclark}

     At $H$ = 26.42 T, the range of the field sweep for the $^{63}$Cu-NMR spin-echo signal covered 0.86 T (from 26.0 T to 26.86 T) at all temperatures with a fixed NMR frequency ($\nu_{0}$ = 298.16 MHz) from the frequency generator, and the sweep used a field step 0.02 T ( i.e., 0.226 MHz in frequency ) for each spin-echo acquisition. The recording for the $^{65}$Cu-NMR spin-echo signals was similar. The field-swept spectra were analyzed with the field-shifted and -summed (FiSS) Fourier transform processing. \cite{wgclark}  

     The corresponding $^{63,65}$Cu-NMR Knight shift and central linewidth at $H$ = 9 T and $H$ = 26.42 T were obtained from the frequency- and field-swept spectra as described above.

     For the purposes of the applied field calibrations, a small piece of thin Al-foil was inserted into the sample coil with the PCCO sample [ note, the $^{27}$Al nucleus in the Al-foil has a gyromagnetic ratio $^{27}$$\gamma$ = 11.0943 MHz/T and an isotropic Knight shift $^{27}$$K$ = 0.162$\%$, i.e., the $^{27}$Al has an effective gyromagnetic ratio $^{27}$$\gamma_{eff}$ = (1 + 0.162$\%$)$\times$11.0943 = 11.112 MHz/T ]. 

     Also we did the DC magnetic susceptibility measurements with an applied magnetic field 3000 Oe upon cooling in temperature from 300 K down to 2 K, using a commercial SQUID magnetometer.
\section{results}
\subsection{$^{63,65}$Cu-NMR spectra}
     Figure 1 shows the $^{63,65}$Cu-NMR spectra with an applied magnetic field $H$ = 9 T at a typical temperature $T$ = 50 K, plotted as the $^{63,65}$Cu-NMR spin-echo amplitude vs frequency shift $\nu$ $-$ $\nu_{rf}$, where $\nu_{\text{rf}}$ is a reference frequency (here $\nu_{\text{rf}}$ = 106 MHz). The area of each spectrum curve (above its baseline) at both $H$ $\parallel$ $c$ and $H$ $\perp$ $c$ is normalized to be 1 for comparison. Theoretically each copper nucleus's spectrum is expected to have a central line plus two symmetric quadrupolar satellites due to the $^{63,65}$Cu spin quantum $m$ = + 1/2 $\leftrightarrow$ $-$ 1/2 (central) and $\pm$ 3/2 $\leftrightarrow$ $\pm$ 1/2 (satellites) transitions, respectively. 

     Instead of sharp satellite spectra, the satellite spectra are extremely broad, with structures that spread across the sharp central lines and overlap between them. The overlap also extends significantly between the two copper isotopes, especially at $H$ $\parallel$ $c$, and their spectra totally cover a range of $\sim$ 18 MHz in frequency. Interestingly, the quadrupolar satellites become narrower and the their peaks become observable (see Fig. 2) upon cooling in temperatures below $\sim$ 20 K. The ratio for the areas below each spectrum curve for each Cu isotope for the satellites and central line in total is $\sim$ 55/45, which is close to the theoretically expected value 60/40. \cite{slichter2}  
\begin{figure}
\includegraphics[scale= 0.36]{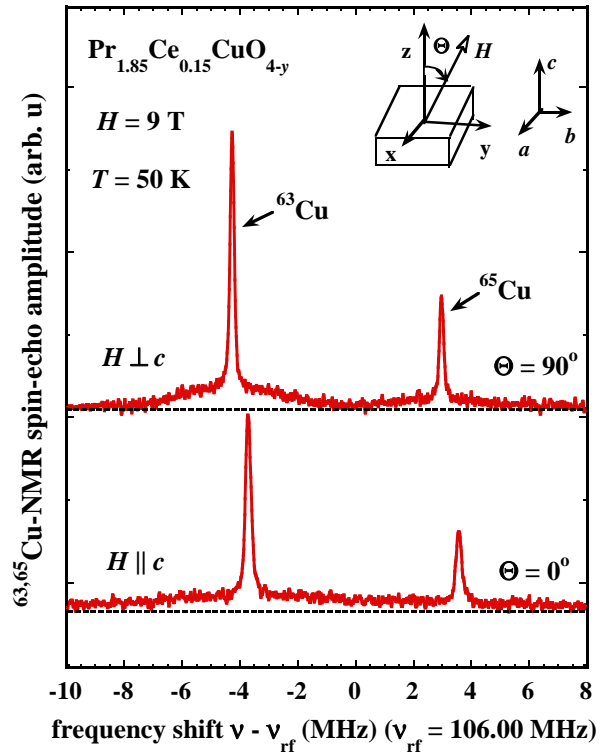}
\caption{(Color online) $^{63,65}$Cu-NMR frequency-swept spectra of a single crystal Pr$_{1.85}$Ce$_{0.15}$CuO$_{4-y}$ (PCCO) at $T$ = 50 K with $H$ = 9 T. The dashed lines are the baselines for the spectra. The upper right corner indicates the lattice axis directions and the direction of $H$ in the $yz$-plane with an angle $\Theta$ relative to the $c$-axis of the thin plate-like crystal sample. The value of $\nu_{\rm{rf}}$ is a reference frequency used for the plot. \label{fig1}}
\end{figure}
\begin{figure}
\includegraphics[scale= 0.32]{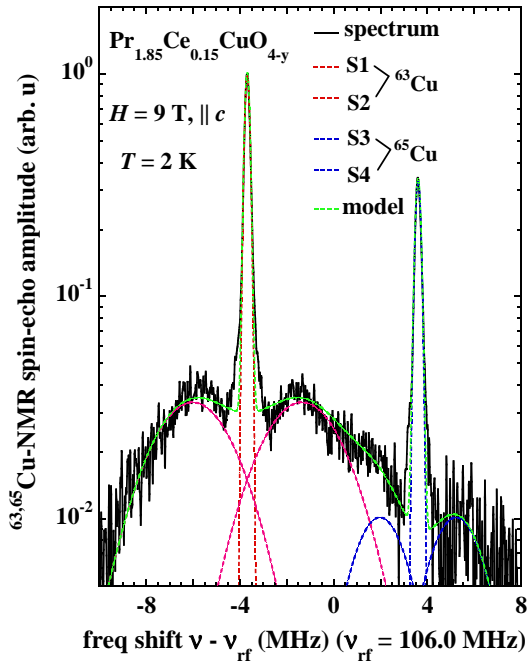}
\caption{(Color online) The fit of the full $^{63,65}$Cu-NMR frequency-swept spectra of a single crystal Pr$_{1.85}$Ce$_{0.15}$CuO$_{4-y}$ (PCCO) at temperature $T$ = 2 K with $H$ = 9 T, $\parallel$ $c$, using a Gaussian model. The dashed red and blue curves are the fit for the satellites of $^{63}$Cu and $^{65}$Cu, respectively, and the dashed green curves are the fit total from the model as compared with the measured spectrum (the solid black curve) (for convenience, the vertical axis is plotted using a logarithmic scale).  \label{fig2}}
\end{figure}

     Theoretically, in the high field limit where the Zeeman splitting (Hamiltonian $H_{\text{Zeeman}}$ = $-\gamma_{I}\hbar\vec{\bf{I}}\cdot\vec{\bf{H}}$) is dominant, for a spin $I$ = 3/2 nucleus the central line has a quadrupolar frequency shift to the 2nd order as \cite{slichter2} $\Delta\nu_{cQ}^{(1)}$ = 0, and
\begin{equation}
\Delta \nu_{cQ}^{(2)} = \frac{3\nu_{Q}^{2}}{16 \nu_{0}} (1 - \cos^{2} \Theta )(1 - 9\cos^{2} \Theta ),
\end{equation} 
while the two satellites have the 1st order quadrupolar frequency shifts \cite{slichter2}
\begin{equation}
\Delta \nu_{sQ} ^{(1)} = \frac{\nu_{Q}}{2}(3\cos^{2} \Theta - 1),
\end{equation}
arising from the electric quadrupole interaction of the nuclear quadrupolar moment($Q$) with the EFG, where $\nu_{Q}$ = $eQV_{zz}/2h$, called the nuclear quadrupolar resonance (NQR) frequency, $h$ is the Planck constant, $e$ is the charge of an electron, and $\nu_{0}$ = $\gamma_{I} H$. The value of $Q$ for $^{63}$Cu is $^{63}Q$ = $-$ 0.211 barns, and for $^{65}$Cu it is $^{65}Q$ = $-$ 0.195 barns (note, 1 barn = 10$^{-28}$ m$^{2}$). Here the principle axes of the EFG ($V_{xx}$, $V_{yy}$, and $V_{zz}$) at the Cu site can be chosen along the lattice $a$, $b$ and $c$ axes, respectively, and then the anisotropic EFG tensor $\eta$ = ($V_{xx} - V_{yy})/V_{zz}$ = 0, due to the tetragonal lattice symmetry.

      With the analysis using Eqs. (1)$-$(2) for the angular dependence of the $^{63, 65}$Cu-NMR spectra, we found $^{63}\nu_{Q}$ = 2.17 $\pm$ 0.03 MHz for the $^{63}$Cu, and $^{65}\nu_{Q}$ = 2.08 $\pm$ 0.04 MHz for the $^{65}$Cu. This gives an experimental ratio of $^{63}\nu_{Q}$/$^{65}\nu_{Q}$ $\approx$ 1.05, which also agrees with the theoretically equivalent ratio of $^{63}Q$/$^{65}Q$ = 1.08. \cite{slichter2}
\begin{figure}
\includegraphics[scale= 0.32]{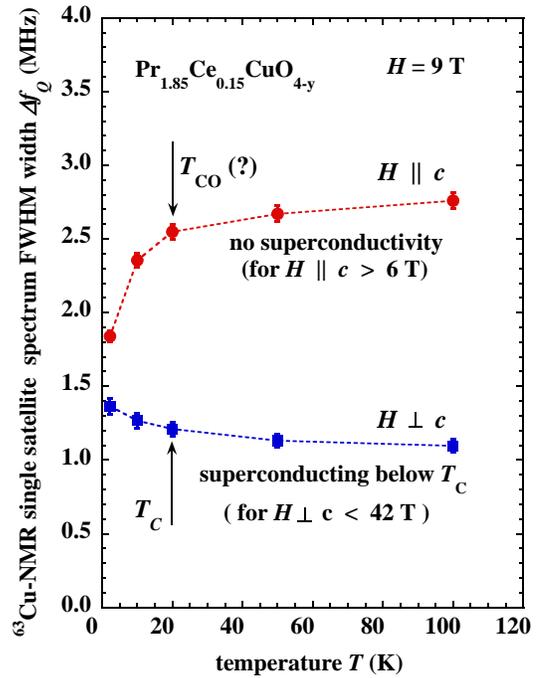}
\caption{(Color online) $T-$dependence of the $^{63}$Cu-NMR single satellite spectrum FWHM width of Pr$_{1.85}$Ce$_{0.15}$CuO$_{4-y}$ (PCCO) at $H$ = 9 T. The downward arrow indicates the temperature where a significant drop of the satellite linewidth starts (labeled as $T_{CO}$ for possible charge ordering) for $H$ $\parallel$ $c$ when the superconductivity is fully suppressed. \label{fig3}} 
\end{figure}
\begin{figure}
\includegraphics[scale= 0.35]{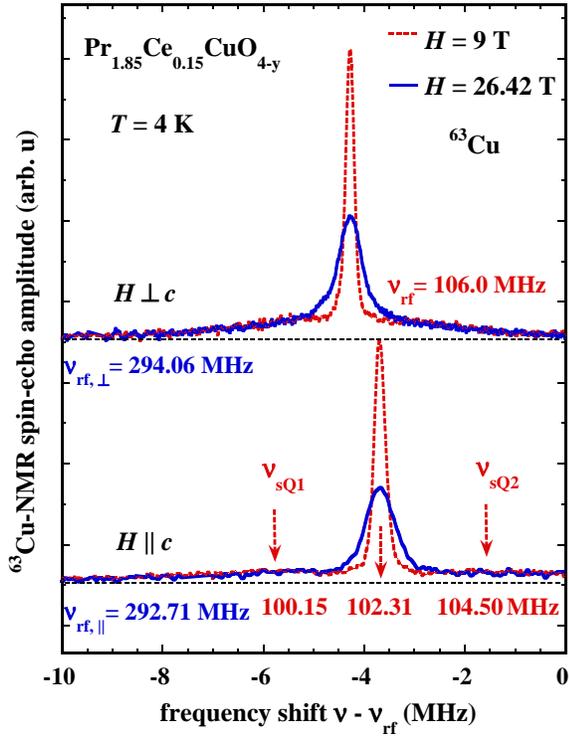}
\caption{(Color online) $^{63}$Cu-NMR frequency-swept spectra at $H$ = 9 T (dashed red curves) and field-swept spectra at $H$ = 26.42 T (solid blue curves) of a single crystal Pr$_{1.85}$Ce$_{0.15}$CuO$_{4-y}$ (PCCO) at $T$ = 4 K. Different values of the reference frequency $\nu_{\rm{rf}}$ are used for the plot. The dashed (red) arrows indicate the satellite peak positions at $H$ $\parallel$ $c$. \label{fig4}}
\end{figure}

      However, this value of $\nu_{Q}$ for the $^{63, 65}$Cu in PCCO is very small, and the $^{63, 65}$Cu-NMR spectrum satellites are extremely broad, in comparison with those in the hole-doped counterparts LSCO ($^{63}\nu_{Q}$ $\sim$ 35 MHz) \cite{kumagai} and YBCO ($^{63}\nu_{Q}$ = 31 MHz), \cite{warren} or its parent compound Pr$_{2}$CuO$_{4-y}$ (PCO) ($^{63}\nu_{Q}$ $\sim$ 15 MHz), \cite{wu} which is an antiferromagnet. Similar observations were also obtained in the electron-doped compound NCCO \cite{abe} which has a value of $^{63}\nu_{Q}$ $\sim$ 1 MHz according to our estimate, while its parent compound Nd$_{2}$CuO$_{4-y}$ (NCO) (also an antiferromagnet) has a value of $^{63}\nu_{Q}$ $\sim$ 14 MHz. \cite{kohori1, abe} Thus the case for PCCO here is rather similar to that in NCCO. 
%

      It is not clear here what is the cause of the very small value of $\nu_{Q}$ (or EFG) and the extremely wide distribution of it at the Cu in PCCO and NCCO, \cite{abe, kohori1} whether they are doping or structure related, for example, even though a similar value for $\nu_{Q}$ was theoretically estimated \cite{zavi} by considering the covalence of Cu and overlapping of the electronic orbitals of Cu and O, as well as the amount of Cu$^{1+}$ impurity associated with the electron doping.  
          
      Figure 2 shows that the full $^{63,65}$Cu-NMR frequency-swept spectra of PCCO can be fitted using a Gaussian model, with high symmetry satellites for each isotope, with the one at temperature $T$ = 2 K and $H$ = 9 T, $\parallel$ $c$, as an example. The fit may not be physical, but it does provide a convenient way of obtaining the satellite width and the peak positions. 

      Figure 3 exhibits the $T-$dependence of the PCCO $^{63}$Cu-NMR single satellite spectrum full-width-half-maximum (FWHM) width at an applied magnetic field $H$ = 9 T. These data indicate a significant narrowing of the satellites at $H$ $\parallel$ $c$, which may indicate a significant charge distribution modification at the Cu site at $T$ $\leq$ $T_{CO}$ =  20 K  where the superconductivity is fully suppressed (with $H$ $\parallel $ $c$ $\geq$ $H_{c2}$ = 6 T), while the corresponding change at $H$ $\perp$ $c$ is negligible when the superconductivity is present or not fully suppressed. This observation may serve as the evidence of the CO as recently found in the electron-doped cuprate NCCO, \cite{ehdsn} while other experimental techniques are still needed for a verification. 

      Figure 4 shows the $^{63}$Cu-NMR spectrum at $H$ = 26.42 T as compared with that at $H$ = 9 T at temperature $T$ = 4 K. The spectra are normalized in area and plotted on top of each other by choosing different values of reference frequency $\nu_{\rm{rf}}$ (note, here $\nu_{\rm{rf}}$ = 106 MHz for both $H$ $\perp$ $c$ and $H$ $\parallel$ $c$ at 9 T). This indicates that the satellites at $H$ = 26.42 T are essentially the same as those at $H$ = 9 T, i.e., no $H$-dependence, and the full width half maximum (FWHM) central linewidth is $\sim$ 3 times wider at $H$ = 26.42 T than at $H$ = 9 T, i.e., $\sim$ proportional to the applied magnetic field $H$. 

      Therefore, this reveals the origin of the internal static electric and magnetic field at the Cu: the satellites are due to the quadrupolar contribution with the charges surrounding the Cu site, while the central transition line is magnetic. 
%
%
\subsection{$^{63}$Cu-NMR Knight shift}
      Figure 5 shows the $T$-dependence of the $^{63}$Cu-NMR Knight shift, $^{63}K(T)$ vs $T$. $^{63}K(T)$ is highly anisotropic and has a fairly weak $T$-dependence at both $H$ $\perp$ $c$ and $H$ $\parallel$ $c$, and there is no significant change of $^{63}K(T)$ across $T_{c}$ (at $H$ $\perp$ $c$) upon cooling. The superconductivity at $H$ $\parallel $ $c$ is completely suppressed by the applied magnetic field $H$ (when $H$ $\geq$ 6 T). \cite{upperc} Moreover, the proportionality of $^{63}K(T)$ in frequency with $H$ further confirms the magnetic origin for the shift of the central line. The Knight shift for $^{65}$Cu [ $^{65}K(T)$ ] (not shown) is the same as $^{63}K(T)$ (for $^{63}$Cu).
\begin{figure}
\includegraphics[scale= 0.36]{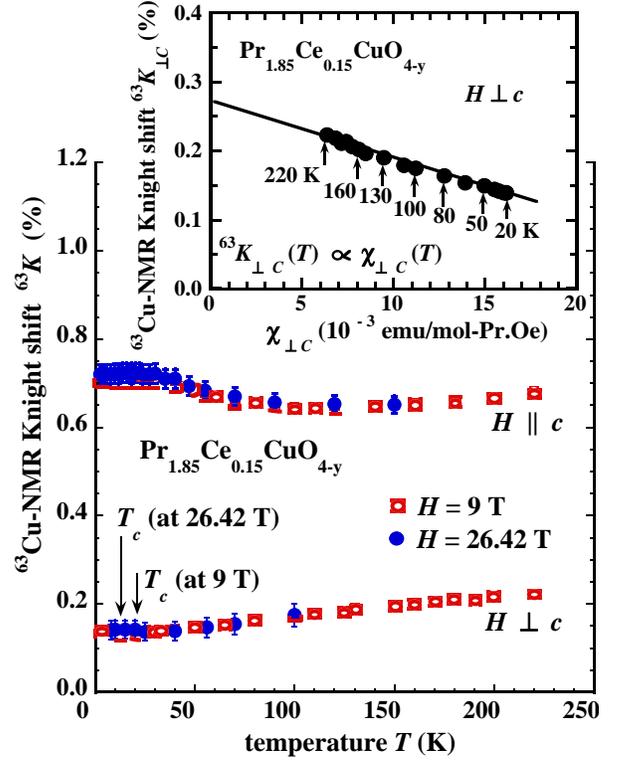}
\caption{(Color online) $T-$dependence of the $^{63}$Cu-NMR Knight shift $^{63}K(T)$ of Pr$_{1.85}$Ce$_{0.15}$CuO$_{4-y}$ (PCCO) at $H$ = 9 T and $H$ = 26.42 T. The downward arrows indicate the temperature $T_{c}$ at the corresponding field for $H$ $\perp$ $c$. The inset shows the linear relation of $^{63}K(T)$ vs $\chi_{\perp c}(T)$ at $H$ $\perp$ $c$, where $\chi_{\perp c}(T)$ is the Pr$^{3+}$ electron paramagnetic susceptibility at $H$ $\perp$ $c$. \label{fig5}} 
\end{figure}

      For the Knight shift $^{63}K(T)$, the internal static magnetic field ($H_{\rm{local}}$) magnitude at the $^{63}$Cu can be written as, $H_{\rm{local}}$ = $H$[1 + $^{63}K(T)$ + $\frac{\Delta \nu_{Q}(\Theta, T)}{\nu_{0}}$], where $\frac{\Delta \nu_{Q}(\Theta, T)}{\nu_{0}}$ = $\frac{\Delta \nu_{cQ} ^{(1)}(\Theta, T) + \Delta \nu_{cQ} ^{(2)}(\Theta, T)}{\nu_{0}}$ = $\frac{3}{16} \frac{\nu_{Q} ^{2}}{\nu_{0} ^{2}}(1 - \cos^{2}\Theta)(1 - 9\cos^{2}\Theta)$. Here $\frac{\Delta \nu_{Q}(\Theta, T)}{\nu_{0}}$ is negligible ($<$ 0.01$\%$) due to the very small value of $\nu_{Q}$, and $^{63}K(T)$ can be expressed as \cite{mehring, sonier}
\begin{eqnarray}
^{63}K(T) & \approx & A_{Cu}^{hf} \chi_{0} + ~(A_{Pr}^{hf} + A_{Pr}^{Dip})~\chi (T) \nonumber \\
                 & &  + ~4\pi(\frac{1}{3} - D)~\frac{\chi_{sample}(T)}{N_{A}\upsilon_{Pr}} + ~^{63}K_{orb},
\end{eqnarray}
where $A_{Cu}^{hf}$ is the anisotropic hyperfine coupling to the Cu$^{2+}$ conduction electron spins in the CuO$_{2}$-plane. $A_{Pr}^{hf}$ and $A_{Pr}^{Dip}$ are the contact hyperfine and dipolar couplings to the Pr$^{3+}$ electron paramagnetic moment, respectively. Here we use $\chi_{0}$ $\approx$ 4 $\times$ 10$^{-5}$ (emu/mol-Cu.Oe), \cite{mattheiss, mehring} which is the static Pauli spin susceptibility of the conduction electrons from the CuO$_{2}$ planes (note, the value of $\chi_{0}$ is very small, and thus we expect the effect to it from the lack of apical oxygen in the PCCO crystal lattice is not significant), $\chi (T)$ is the Pr$^{3+}$ electron paramagnetic susceptibility, and $\chi_{sample}(T)$ is the sample magnetic susceptibility [$\chi_{sample}(T)$ $\approx$ $\chi (T)$ + $\chi_{0}$]. The 3rd term in Eq. (3) is the correction due to bulk demagnetization and Lorentz fields \cite{slichter2}, $\upsilon_{Pr}$ is the unit cell volume/Pr$^{3+}$, $D$ $\approx$ 0.04 and 0.93 at $H$ $\perp$ $c$ and $H$ $\parallel$ $c$, respectively, due to the sample size from our estimate, \cite{gwu} and the last term $^{63}K_{orb}$ comes from the $T-$independent Cu$^{2+}$ 3$d$-orbital contribution. 

      From the PCCO lattice structure we calculated \cite{sebrown} that $A_{Pr,\parallel c}^{Dip}$ = + 6.2 kG/$\mu_{B}$ (or 0.124 T), and $A_{Pr,\perp c}^{Dip}$ = $-$ 3.1 kG/$\mu_{B}$ (or $-$ 0.062 T) (note, 1 T = 5 kG/$\mu_{B}$). We also estimated that $A_{Cu, \parallel c}^{hf}$ = $-$ 100 kG/$\mu_{B}$ (or $-$ 20 T), and $A_{Cu, \perp c}^{hf}$ = + 180 kG/$\mu_{B}$ (or + 36 T), with the consideration of the measured normal state $^{63}$Cu-NMR spin-lattice relaxation. \cite{gwu, sebrown}  
 
      Thus, with the fits to Eq. (3) and the analysis as that shown by the solid line in the inset of Fig. 4 for the Knight shift at $H$ $\perp$ $c$, we obtained $^{63}K_{orb,\perp c}$ = (0.18 $\pm$ 0.01)$\%$, and $A_{Pr,\perp c}^{hf}$ = ($-$ 4.25 $\pm$ 0.1) kG/$\mu_{B}$ [ or $-$(0.85 $\pm$ 0.02) (T) ]. Similarly, for $H$ $\parallel$ $c$, we have $^{63}K_{orb, \parallel c}$ = (0.84 $\pm$ 0.01)$\%$, and $A_{Pr,\parallel c}^{hf}$ = ($-$ 4.20 $\pm$ 0.2) kG/$\mu_{B}$ [ or $-$(0.85 $\pm$ 0.04) (T) ] (here the subscripts / superscripts of $\parallel$ $c$ and $\perp$ $c$ denote the $H$ direction relative to the lattice $c-$axis). This give a high anisotropy ratio of $^{63}K_{orb, \parallel c}$ / $^{63}K_{orb, \perp c}$ = 4.6 $\pm$ 0.1.
       
      Therefore, this indicates that the $^{63}$Cu-NMR Knight shift is dominated by the $T-$independent anisotropic orbital shifts, $^{63}K_{orb, \parallel c}$ (at $H$ $\parallel$ $c$) and $^{63}K_{orb, \perp c}$ (at $H$ $\perp$ $c$), arising from the hyperfine to the Cu$^{2+}$ 3$d$-orbitals, while the weak $T-$dependence of the Knight shift is determined by the isotropic contact hyperfine coupling to the Pr$^{3+}$ paramagnetic spins ($A_{Pr,\parallel c}^{hf}$ = $A_{Pr,\perp c}^{hf}$ $\approx$ $-$ 4.2 kG/$\mu_{B}$ from above), to which the dipolar coupling ($A_{Pr, \parallel c}^{Dip}$ and $A_{Pr,\perp c}^{Dip}$) is only $\sim$ 14.5$\%$ and 7.2$\%$ at $H$ $\parallel$ $c$ and $H$ $\perp$ $c$, respectively. The negative value of $A_{Pr,\parallel c}^{hf}$ ($A_{Pr,\perp c}^{hf}$) $<$ 0, indicates an antiferromagnetic character for the coupling.     
\begin{figure}
\includegraphics[scale= 0.35]{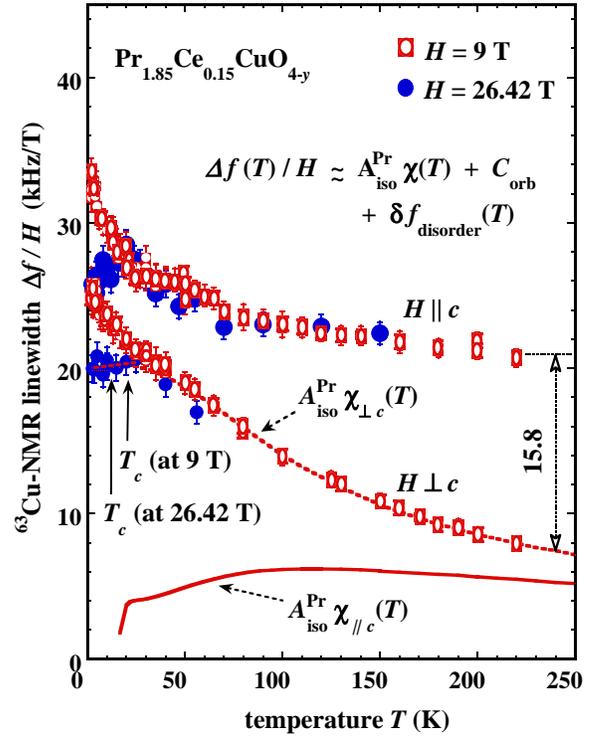}
\caption{(Color online) $T-$dependence of the $^{63}$Cu-NMR central linewidth (FWHM) divided by $H$, $\Delta f(T)$/$H$, of Pr$_{1.85}$Ce$_{0.15}$CuO$_{4-y}$ (PCCO) at $H$ = 9 T and $H$ = 26.42 T. The dashed (at $H$ $\perp$ $c$) and solid (at $H$ $\parallel$ $c$) lines indicate the contribution from the anisotropic Pr$^{3+}$ magnetic susceptibility. \label{fig6}} 
\end{figure}
\subsection{$^{63}$Cu-NMR linewidth}
      Figure 6 shows the $T-$dependence of the $^{63}$Cu-NMR central linewidth (FWHM), $\Delta f(T)$, plotted as $\Delta f(T)$/$H$ in units kHz/T vs $T$. As described above, the central line has a magnetic origin, and it is inhomogeneously broadened upon cooling in temperature. Thus considering the $^{63}$Cu-NMR Knight shift, in which the anisotropic hyperfine from the dipolar field of the Pr$^{3+}$ paramagnetic spins and from the Cu$^{2+}$ conduction electron spins (Pauli) are not significant, we expect the central linewidth to be written as 
\begin{equation}
\Delta f(T)/H \approx A_{\rm{iso}}^{Pr}~\chi (T) + ~C_{\rm{orb}} + ~\delta f_{\rm{disorder}}(T), \\
\end{equation}
where $A_{\rm{iso}}^{Pr}$ is an isotropic constant, $C_{\rm{orb}}$ is the $T$-independent anisotropic Cu$^{2+}$ 3$d$-orbital contribution, and $\delta f_{\rm{disorder}}(T)$ is due to magnetic disorder (if any).
%

       As shown by the solid and dashed lines in Fig. 4, the linewidth data can be well-fitted with Eq. (4) as
\begin{eqnarray}
\Delta f_{\parallel c}(T)/H & \approx & A_{\rm{iso}}^{Pr}~\chi_{|| c}(T) + C_{\rm{orb}, || c}, (T \geq 100 ~\text{K}) \\
\Delta f_{\perp c}(T)/H & \approx & A_{\rm{iso}}^{Pr}~\chi_{\perp c}(T), ~~~~~~~~~~~~(T \geq 20 ~\text{K})
\end{eqnarray}
where the fitted values of $A_{\rm{iso}}^{Pr}$ $\approx$ 1.17 $\times$ 10$^{3}$ [(kHz/T).(mol.Pr.Oe/emu)], $C_{orb, || c}$ $\sim$ 15.8 kHz/T, $C_{\rm{orb}, \perp c}$ $\sim$ 0, $\delta f_{\rm{disorder}}^{~|| c}(T)$ $\sim$ 0, and $\delta f_{\rm{disorder}}^{~\perp c}(T)$ $\sim$ 0, in the temperature range specified above in Eqs. (5)-(6).

       Thus Fig. 6 [ Eqs. (4) - (5) ] reveals that, 1) the $^{63}$Cu-NMR central linewidth $\Delta f(T)$ is essentially proportional to the applied magnetic field $H$ (because of its magnetic origin) except for the development of possible magnetic disorder at low-$T$ ($<$ $\sim$ 25 K), 2) at $H$ $\parallel$ $c$ the linewidth is dominated by the anisotropic Cu$^{2+}$ 3$d$-orbital contributions (due to the Cu$^{2+}$ orbital moments), and 3) at $H$ $\perp$ $c$ the linewidth is almost completely determined by the isotropic contact hyperfine coupling to the Pr$^{3+}$ paramagnetic moments, i.e., there is a negligible contribution from the Cu$^{2+}$ 3$d$-orbital to the internal static magnetic field distribution at the Cu at $H$ $\perp$ $c$. But the Cu$^{2+}$ 3d-orbital contribution always dominates the $^{63}$Cu-NMR Knight shift (internal static magnetic field magnitude) at both $H$ $\perp$ $c$ and $H$ $\parallel$ $c$. This local magnetic field environment at the Cu in PCCO is very different from that in the hole-doped cuprate HTSCs, \cite{mehring} where effects from ions of large spin paramagnetic moment (like Pr$^{3+}$ in PCCO or Nd$^{3+}$ in NCCO) do not exist. 

       The origin for the development of possible magnetic disorder at low $T$ (seen from the linewidth) is not clear, even though it could come from minor impurity oxygen \cite{miyagawa} trapped in the sample during the sample synthesis process.  
\subsection{$^{63}$Cu 3d-orbital energy splitting}
       Finally, the parameters for the energy splitting of the Cu$^{2+}$ 3$d$ orbitals in the CuO$_{2}$-plane can be obtained \cite{pennington, barrett} through the orbital Knight shift anisotropy as \cite{barrett, mcMahan}  
\begin{eqnarray}
^{63}K_{\rm{orb},\parallel c}/^{63}K_{\rm{orb},\perp c} & = & 4 (E_{xz} - E_{x^{2} - y^{2}})/(E_{xy} - E_{x^{2} - y^{2}}),\nonumber \\
E_{xz} & = & E_{yz},
\end{eqnarray}
where $E_{xy}$, $E_{xz}$, $E_{yz}$, and $E_{x^{2} - y^{2}}$ are the energy levels of the Cu$^{2+}$ d$_{xy}$, d$_{xz}$, d$_{yz}$ and d$_{x^{2} - y^{2}}$ orbitals, respectively. 

      By using the obtained anisotropy ratio of $^{63}K_{orb, \parallel c}$/$^{63}K_{orb, \perp c}$ = 4.6 $\pm$ 0.1, we have the energy state of the Cu$^{2+}$ 3$d$-electrons as $\frac{E_{xz} - E_{x^{2} - y^{2}}}{E_{xy} - E_{x^{2} - y^{2}}} = 1.15 \pm 0.01$, i.e., $E_{xz} = E_{yz} > E_{xy}$. This agrees well with the theoretical calculation \cite{mcMahan} and observation \cite{jaleiro} regarding the Cu$^{2+}$ 3$d$-orbital energy levels of $E_{xz}$, $E_{yz}$ and $E_{xy}$ relative to the ground level $E_{x^{2} - y^{2}}$, suggesting a similar high anisotropy of the Cu$^{2+}$ 3$d$ orbital shift and similar electronic energy state of the Cu$^{2+}$ electron itself in the CuO$_{2}$-plane to those in the hole-doped cuprate HTSCs. \cite{jaleiro}  
\section{Conclusion}
      In summary, a very small NQR frequency $\nu_{Q}$ $\sim$ 2.2 MHz is obtained with the observation of an unusual $^{63, 65}$Cu-NMR spectrum, which shows a very small electric field gradient (EFG) and an extremely wide continuous distribution of it ($\Delta \nu_{Q}$ $\sim$ 18 MHz) at the copper in PCCO. Upon cooling in temperature, the distribution of EFG becomes significantly narrower below 20 K at $H$ $\parallel$ $c$ where the superconductivity is completely suppressed, indicating a significant change in the charge distribution modulation at the Cu site. Other experimental techniques are needed to verify whether this is due to CO or a different type of charge distribution modulation.

      The $^{63,65}$Cu-NMR Knight shift and the central linewidth are proportional to the externally applied magnetic field, with an orbital shift anisotropy of $\sim$ 4.6. We find that the magnitude of the internal static magnetic field at the copper at both $H$ $\perp$ $c$ and $H$ $\parallel$ $c$ is dominated by the $T$-independent anisotropic hyperfine coupling to the Cu$^{2+}$ 3$d$ orbitals, while its weak $T$-dependence is mainly determined by the isotropic contact hyperfine coupling to the paramagnetic Pr$^{3+}$ electron spins, which is also responsible for the full distribution of the internal static magnetic field at the copper at $H$ $\perp$ $c$. But at $H$ $\parallel$ $c$, the distribution of the internal static magnetic field at the copper is dominated by the Cu$^{2+}$ 3$d$-orbital contributions through anisotropic hyperfine couplings. Thus, unlike the Cu$^{2+}$ 3d orbitals, the Cu$^{2+}$ spins provide a small contribution to the internal static magnetic field at the copper in PCCO. This unusual internal static electric and magnetic field environment at the copper in cuprate HTSCs may provide new insight into the understanding of the high-$T_{c}$ superconductivity.    
\begin{acknowledgments}
      The work at NHMFL was supported by NSF under Cooperative Agreement No. DMR-0654118 and the State of Florida, at UM by DMR-1104256 (RLG), at University of West Florida by SCA/2009-2012 (G. Wu) and at UCLA by NSF Grants DMR-0334869 (WGC). We thank Stuart E. Brown for helpful discussions and support.
\end{acknowledgments}


\begin{references}
\bibitem{upperc} Our upper critical field $H_{c2}$ vs $T$ measurement shows $H_{c2, \perp c}$(0)/$H_{c2, \parallel c}$(0) = 42 T/6 T, where $H_{c2, \perp c}$(0) and $H_{c2, \parallel c}$(0) are $H_{c2}$ values at $H$ $\perp$ $c$ and $H$ $\parallel$ $c$ when $T$ $\rightarrow$ 0, respectively. At $\perp$ $c$, $T_{c}$ = 20 K and 12 K for $H$ = 9 T and 26.42 T $\perp$ $c$, respectively.
\end{references}
\end{document}